# Dot-Tracking Methodology for Background Oriented Schlieren (BOS)

Lalit K. Rajendran[1], Sally P. M. Bane[1] and Pavlos P. Vlachos[2]
1: School of Aeronautics and Astronautics, Purdue University, USA
2: School of Mechanical Engineering, Purdue University, USA

## 1   Abstract

We propose a dot-tracking methodology for processing Background Oriented Schlieren (BOS) images. The method significantly improves the accuracy, precision and spatial resolution compared to conventional cross-correlation algorithms. Our methodology utilizes the prior information about the dot pattern such as the location, size and number of dots to provide near 100% yield even for high dot densities (20 dots/32x32 pix.) and is robust to image noise. We also propose an improvement to the displacement estimation step in the tracking process, especially for noisy images, using a "correlation correction", whereby we combine the spatial resolution benefit of the tracking method and the smoothing property of the correlation method to increase the dynamic range of the overall measurement process. We evaluate the performance of the method with synthetic BOS images of buoyancy driven turbulence rendered using ray tracing simulations, and experimental images of flow in the exit plane of a converging-diverging nozzle.

## 2   Introduction

Background Oriented Schlieren (BOS) is an optical flow diagnostic technique used to measure density gradients in a flow field by tracking the apparent distortion of a target dot pattern. Since density and refractive index are proportional for fluids, density gradients in a flow are associated with refractive index gradients, and an object viewed through a variable density medium will appear distorted due to the refraction of light rays traversing the medium. The distortion of the dot pattern is typically estimated by cross-correlating an image of the dot pattern without the density gradients (called the reference image) with a distorted image viewed through the density gradients (called the gradient image) using techniques borrowed from Particle Image Velocimetry (PIV) [1]–[4]. Alternatively, the distortion can also be estimated using optical flow algorithms [5].

Low spatial resolution has been traditionally one of the limitations of BOS compared to the traditional schlieren technique, [6], [7] and is due to the large interrogation window sizes required for the PIV-type cross-correlation algorithms to ensure sufficient signal to noise ratio for the measurements [8], [9]. While multi-pass interrogation schemes and window overlap can increase the spatial resolution [10]–[12], adjacent vectors still have some spatial dependence and do not constitute purely independent measurements.

An alternative processing approach that can increase the spatial resolution is tracking individual dots from one image to the next, as done in Particle Tracking Velocimetry (PTV) applications [13]–[15]. Despite the popularity of PTV methods, such analysis has not received attention for BOS images. The primary factor controlling the performance of PTV methods is the ratio of particle displacement across images to inter-particle distance in the same image, because it affects the reliability of matching the same particle between the two frames. Since typical displacements in PTV applications are about 10 pixels, they are traditionally limited to low seeding densities. However, the displacements are typically very low in BOS applications (< 2-3 pixels in most cases), so large dot densities can be used before the accuracy of the dot matching procedure is affected. For example, even with 20 dots in a 32x32 pixel window, the inter-dot distance is still about 3-4 pixels if a dot is about 3 pixels in diameter, so the ratio of dot displacement to inter dot distance is low enough to ensure reliable measurements. Perhaps more importantly, the dot patterns used for BOS experiments are manufactured, and hence all the information about the dots such as their location, size and number are known. Therefore, the tracking method can be applied in an iterative manner till all the dots in the frame have been tracked, to achieve near 100% vector yield. This has also been noted by Charruault et. al. [16] who proposed a tracking algorithm for BOS based on Voronoi cells, and showed that their tracking approach can measure much larger image deformations compared to correlation when applied to an air-cavity interface. Thus, dot tracking algorithms are especially well suited for BOS, and we will show in the following sections that they can also substantially increase the accuracy and spatial resolution of the measurements.

In the following sections, we will introduce a dot tracking methodology for BOS, and compare its performance with the traditional cross-correlation method using synthetic images rendered with known density fields and experimental images of the flow field in the exit plane of a converging-diverging nozzle.

## 3 Dot Tracking Methodology

A schematic of the dot tracking methodology is shown in Figure 1. We first describe the standard tracking method, which consists of three steps, (i) particle identification, (ii) sizing and centroid estimation and (iii) tracking.

In standard tracking applications, the particles/dots in the image are isolated from the background using intensity thresholding and segmentation procedures. This can be done using a static intensity threshold or a dynamic threshold using a dilatation-erosion procedure, where the threshold is systematically varied to identify overlapping dots. [14] [17] The main limitation common to all these methods is the choice of the intensity threshold which can either lead to missed particles/dots if the threshold is high or falsely identified particles/dots if the threshold is low. This becomes especially problematic in cases with varying background illumination where the same threshold could be "high" in one part of the image with low illumination, and "low" in other parts of the image with higher background illumination. It also makes the method more error prone in the presence of image noise, due to noisy pixels being falsely identified as particles/dots.

Next in the sizing step, the geometrical properties (centroid and diameter) of the identified particles/dots are estimated to sub-pixel resolution. This can be accomplished using a variety of schemes ranging from a geometrical/intensity-weighted centroid to Gaussian sub-pixel fitting schemes such as the Three/Four Point Gaussian fits and the Least Square Gaussian fit. [18]

Finally in the tracking step, for each dot in the first frame, its corresponding match in the second frame is estimated using a nearest neighbor algorithm. While the nearest neighbor is typically defined as the dot in the second frame that lies closest to the estimated location of the dot in the first frame, it can be generalized using a multi-parametric approach where other properties of the dot such as the peak intensity and diameter can be included to define a weighted residual. The dot in the second frame having the lowest weighted residual and within a pre-defined search radius is defined as the match of the given dot in the first frame, and the dot displacement between the two frames is calculated. [14]

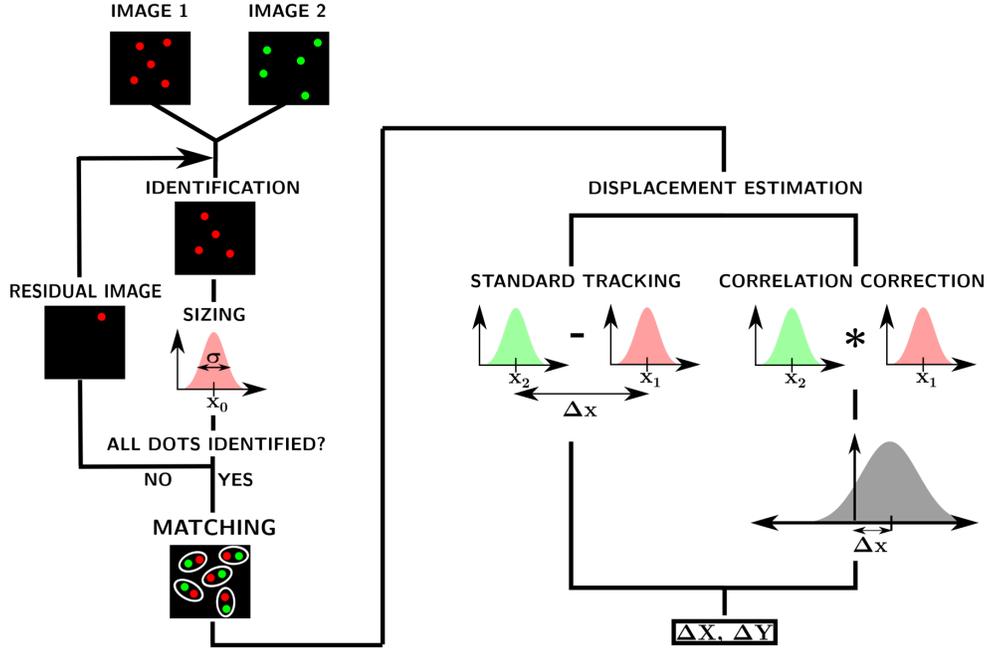

**Figure 1.** Dot Tracking Methodology.

The primary novel contribution of this work is to recognize and utilize in an optimal fashion, the prior information about the dot pattern that is available from the target fabrication, and use this information to improve the overall accuracy and robustness of the method. In the identification step, instead of choosing an intensity threshold to separate from the dots from the background, we use the known location of the dots on the target, and the mapping function of the camera (obtained from calibration) to project the dot locations on the image plane and create a window around this location. The mapping function of the camera can be determined using a calibration process, and a polynomial mapping function as proposed by Soloff et. al. is used in this work. [19] The size of the window is chosen to be slightly larger than the diameter of the dot, where the dot diameter can either be specified beforehand based on the manufacturing details or can be calculated from the diameter of the cross-correlation peak ($d_p = d_{cc}/\sqrt{2}$). [20]

This window will contain pixels corresponding to the true dot as well as noisy pixels. To separate the dot from the noisy pixels, we use the dynamic segmentation procedure based on erosion-dilatation proposed by Cardwell et. al. [14] to segment the window of pixels to create pixel blobs. In cases where more than pixel blob is detected, we sort the pixel blobs based on their pixel area, peak intensity and distance of the peak from the predicted dot location. We calculate a weighted average of the three properties defined as,

$$C_p = \frac{\left(W_A * \left(\frac{A_p}{max(A_p)}\right) + W_I * \left(\frac{I_p}{max(I_p)}\right) + W_{\Delta x} * \left(1 - \frac{\Delta x_p}{max(\Delta x_p)}\right)\right)}{W_A + W_I + W_{\Delta x}} \quad (1)$$

where $A_p, I_p, \Delta x_p$ are the pixel area, peak intensity and distance respectively for the $p^{th}$ blob, and $W_A, W_I, W_{\Delta x}$ are the associated weights. The pixel blob with the highest weighed average is considered to the true dot and the pixels corresponding to the other blobs are set to zero. For the analysis reported in this paper, the weights were set to 1/3 (equally weighted), but these can be changed for other situations. Once the pixel map for the dot has been extracted, a centroid estimation procedure is performed based on subpixel fitting. An example of this procedure is shown in Figure 2.

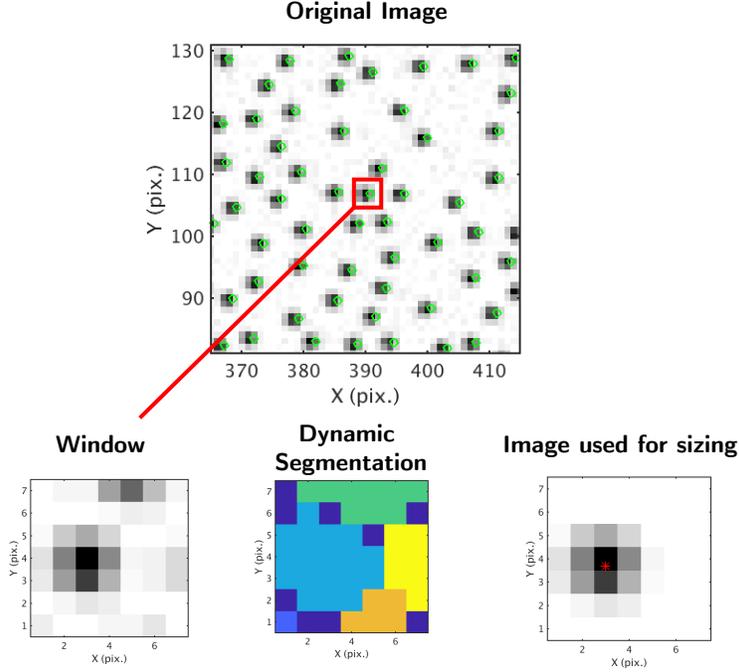

**Figure 2.** Illustration of the dot identification step using prior information about the dot location.

While it is straightforward to see that this approach will work for the reference image (without density gradients), it will also work for the gradient image (with density gradients), because the dot displacements are generally very small (< 2 pix.). Hence the actual location of the dot in the second frame will still be quite close to the predicted location, and since the window is taken to be larger than the dot diameter, it will be large enough to enclose the dot in the second frame as well.

Further, the identification and sizing steps can be performed in an iterative manner to ensure that all the dots on the target have been located. This is done by creating a residual image at the end of each iteration by removing the intensity contribution from the identified dots, as shown in Figure 1. The intensity of the residual image is given by,

$$I^{k+1} = I^k - \sum_{p=1}^{N_p^k} I_{0,p} \exp\left[-\left\{\frac{(X-X_p)^2 + (Y-Y_p)^2}{2\eta_p^2}\right\}\right] \quad (2)$$

where $I^k$ is the image intensity after $k$ iterations, $p$ is the dot index, $N_p^k$ is the number of dots identified in the $k^{th}$ iteration, and $X_p, Y_p, \eta_p$ and $I_{0,p}$ are the positions, diameter and the peak intensity of the $p^{th}$ identified dot. In this way, we are able to improve the accuracy of the method by avoiding incorrect matches and displacement errors due to failed identifications.

We also propose an improvement to the displacement estimation step after the dot matching procedure. Traditionally the displacement is estimated by subtracting the centroids of the two matched particles/dots, but this is error prone because the subpixel fitting procedure is highly sensitive to noise leading to a large position error. This will in turn lead to increased errors in the calculation of the displacements, density gradients and as well as the density field from 2D integration of the density gradients. Further, since the displacements in BOS experiments are typically low, this also severely limits the dynamic range of the measurement. To alleviate this problem, we perform a correlation of the intensity maps of the dots in the two frames to estimate the displacement, as the noise in the pixels is expected to be uncorrelated between the two frames. The intensity maps used are the ones obtained at the end of the identification process where the pixels corresponding to noise/other peaks have been zeroed out, to further improve the correlation signal to noise ratio. In addition, a minimum subtraction operation is also performed where the minimum intensity is taken from the dot window prior to zeroing out the noisy pixels. We refer to this step as a "correlation correction" and in this way we are able to combine the spatial resolution benefit of the tracking method with the noise robustness of the correlation method. This step is illustrated in Figure 3.

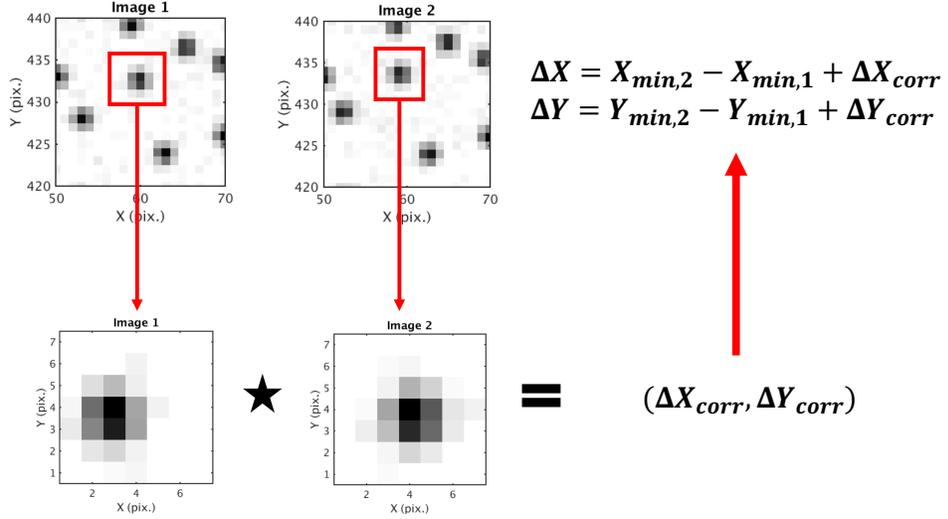

**Figure 3.** Displacement estimation by correlating the intensity maps of the two matched dots.

In the following sections, we will apply this tracking methodology to both synthetic and experimental BOS images and show a substantial improvement in the accuracy, precision and spatial resolution of the results.

## 4 Error Analysis with Synthetic Images

To provide a baseline for comparing the performance of the correlation and tracking methods, an error analysis was first performed using synthetic images rendered with density fields obtained from Direct Numerical Simulation (DNS) data of homogeneous buoyancy driven turbulence performed by Livescu et. al., [21], [22] and available at the Johns Hopkins turbulence database. [23], [24]. The flow involves sharp changes in density over small spatial regions, and hence provides a suitable test case for assessing the spatial resolution of the processing schemes.

### 4.1 Image Generation Methodology

The synthetic BOS images are rendered using a ray tracing-based image generation methodology described in more detail in Rajendran et. al. (2018). [25] The BOS experiment is simulated by generating light rays from the dot pattern and traced through the density gradient field and optical elements to the camera sensor. The trajectory of the light rays through the density gradient field is calculated by solving Fermat's equation:

$$\frac{d}{d\xi}\left(n\frac{d\vec{x}}{d\xi}\right) = \nabla n \qquad (3)$$

using a 4[th] order Runge-Kutta algorithm following established methods in gradient-index optics literature.[26], [27] The refraction through the lens is modelled by Snell's law and the diffraction pattern on the image sensor is modeled using a Gaussian distribution as in synthetic PIV image generation. [28], [29] The computationally intensive ray tracing process is parallelized using Graphics Processing Units (GPUs) and images rendered using this methodology display real world features such as blurring and optical aberrations which can be adjusted in a controlled manner. This methodology has been tested and validated using known density fields. [25] At the end of the ray tracing simulations, the final light ray deflections on the camera sensor for all rays originating from a dot are averaged and used as ground truth for displacement of that dot. This process is repeated for all dots on the pattern to estimate the true displacements throughout the field of view.

Two dimensional (x,y) slices of the flow field from five time instants were chosen, and for each time instant, a three-dimensional density volume was constructed by stacking the same two-dimensional slice along the z-direction, thereby ensuring that the gradient of density in the z direction was zero. This was done to account for the depth integration limitation of BOS measurements and decouple it from the error analysis. Further, the density data was multiplied by 1.225 kg/m³ to simulate air and enclosed in a three-dimensional volume of size 32 mm x 32 mm x 10 mm.

Images of the density field at these snapshots are shown in Figure 4, along with the density gradient, the theoretical light ray displacements and the light ray displacements from the ray tracing simulations. The theoretical displacements were calculated by

$$\Delta \vec{X} = \frac{MZ_D}{n_0} \int_{z_i}^{z_f} \nabla n \, dz \\ \approx \frac{MZ_D K}{n_0} (\nabla \rho)_{avg} \Delta z \quad (4)$$

where $\Delta \vec{X}$ is the theoretical deflection of a light ray, $(\nabla \rho)_{avg}$ is the path-averaged value of the density gradient, $K$ is the Gladstone-Dale constant, $n_0$ is the ambient refractive index, and $\Delta z$ is the thickness of the density gradient field. [1] For the present simulations, the values of the parameters were $M = 0.12$, $Z_D = 0.25$ m, and $\Delta z = 10$ mm.

The final light ray locations will be randomly scattered on the image plane corresponding to the image of the dots. The deflections at these random locations are interpolated to a regular grid for displaying the figures shown in Figure 4, where the contours of simulated light ray deflections are seen to correspond reasonably well to the theoretical displacements, and in both cases the regions of large displacements correspond to regions of large density gradients. The simulated light ray deflections will not match the theoretical displacements exactly, partly due to 1) small angle approximations used in the theory and 2) the spatial resolution limitation of the experimental setup due to the finite angle of a ray cone emerging from the target. Both of these are well known characteristics of BOS experiments. [6] However, these features are common to the images processed by both the correlation and tracking algorithms, and these simulated light ray deflections are considered as the ground truth for conducting the error analysis.

For the present simulations, the dot diameter was 3 pixels and the dot density was 20 dots per 32x32 pixel window. As dot patterns can be manufactured in a controlled manner for BOS experiments, we use dot patterns without overlapping dots. For each snapshot of the DNS, ten image pairs were rendered, and the images were corrupted with zero-mean Gaussian noise with a standard deviation of 1, 3 and 5% of the peak image intensity.

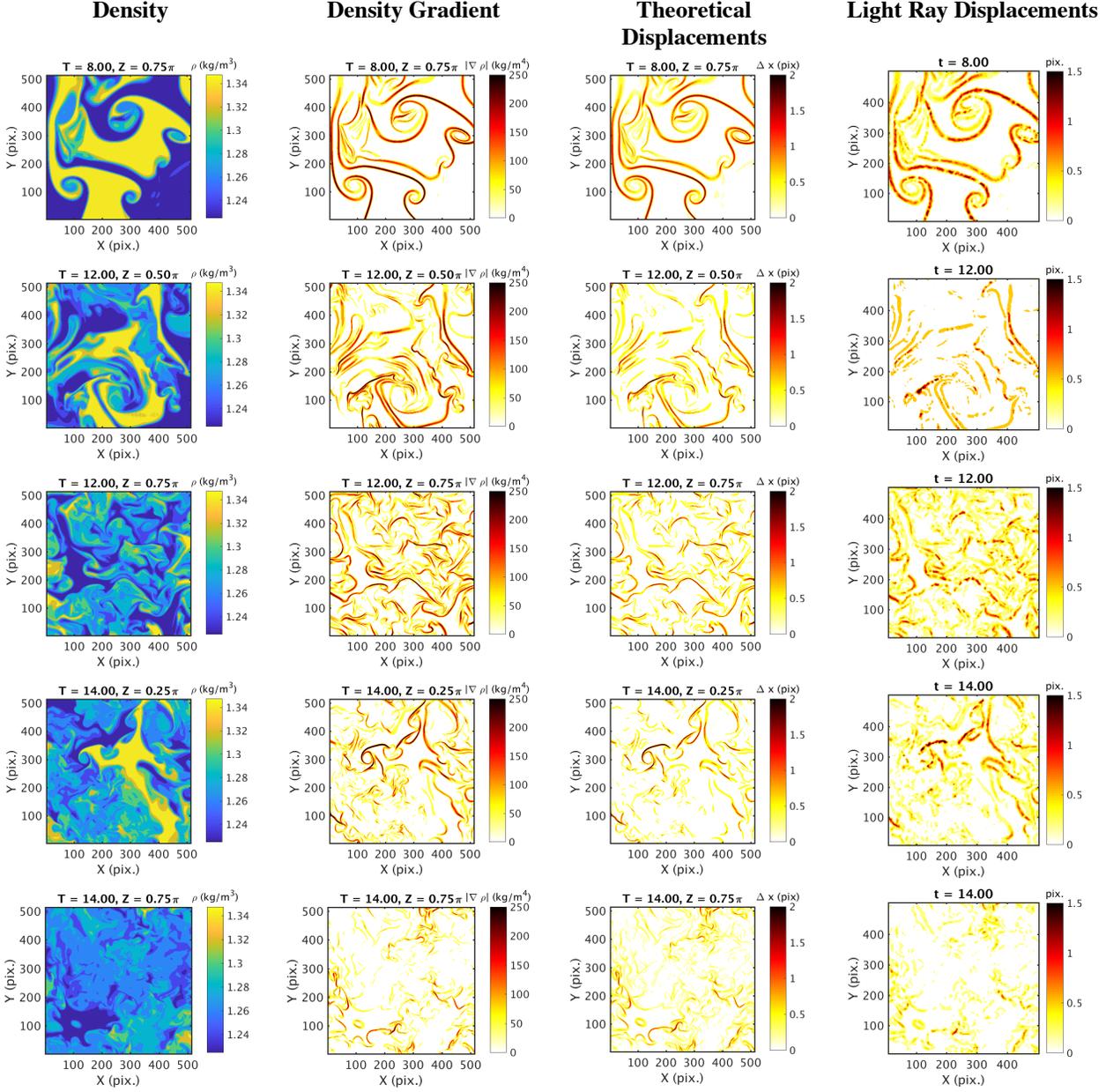

**Figure 4.** Contours of density, density gradients, theoretical displacements and simulated light ray displacements for the five snapshots of DNS data used in the error analysis.

## 4.2 Results

The images were processed using both traditional cross-correlation and the dot tracking method described in Section 3. For the correlation, a multi-grid window deformation method was used with a window size of 32x32 followed by 16x16 pix. without window overlap. For the tracking method a three-point Gaussian subpixel fit was used both for centroid estimation as well as for the displacement estimation using the correlation correction. Then the errors in the final displacements were calculated using the light ray deflection from the ray tracing as the ground truth. Further, the errors were divided into two groups depending on whether the true displacement in that region was above or below a certain threshold. This was done to differentiate the errors due to background image noise, from errors due to lack of spatial resolution. The threshold was chosen to be half the standard deviation of the histogram of *theoretical displacements*. For each noise level, above 500,000 vectors were used in calculating the error distribution, to ensure statistical convergence of the results.

The CDF of the error distribution is shown in Figure 5 for all the noise levels. For the case with zero noise, both tracking methods far outperform the correlation method, where nearly all the vectors have an error below 0.01 pixels as opposed to the correlation algorithm, where the error level corresponding to 90% of the vectors is over 0.1 pixels, which is an order of magnitude more than the tracking. Further, the error levels for the correlation are seen to be higher for vectors above the displacement threshold, as these lie in regions with sharp displacement gradients that cannot be captured by the correlation algorithm.

As the noise level increases, the errors for the tracking methods are seen to increase while they remain nearly the same for the correlation method. The main contribution to error in the tracking method is the position error from the centroid estimation, which is sensitive to image noise. However, the correlation method is robust to image noise in general, because the pixel noise across the two frames will be uncorrelated and hence have a lesser effect on the signal to noise ratio of the correlation plane.

The performance of the processing algorithms can be further understood from looking at the error distributions for the vectors above and below the threshold separately. For the higher noise levels, it is seen that the error from tracking approaches the error from correlation for vectors below the threshold, as the error in this region is dominated by position error due to image noise. However, the tracking methods still outperform the correlation method for vectors above the threshold as the error in this region is dominated by spatial resolution requirements due to sharp displacement gradients in the flow field. The tracking methods are seen to be robust to this effect, with nearly the same noise levels for vectors both above and below the threshold.

Finally, it is seen that the tracking method with the correlation correction performs best even for the highest image noise as it combines the spatial resolution benefit of tracking and the smoothing effect of the correlation. This is particularly evident from Figure 5 (g) where it is seen that for vectors below the threshold, the noise level is so high that the pure tracking method performs poorer than the correlation, however tracking with correlation correction still maintains the same error level as full correlation. For vectors above the threshold the tracking method with correlation correction is still able to maintain the high spatial resolution and performs best overall.

Also shown are the errors in the estimates of the gradient of displacements, corresponding to the second derivative of density. This quantity is needed to perform 2D integration of the density gradient field by solving the Poisson equation, which requires the calculation of the Laplacian of the density field. [31] Again, the dot tracking methods far outperform traditional cross-correlation for all noise levels both above and below the threshold, possibly because the displacement gradient is even more sensitive to the spatial resolution of the schemes.

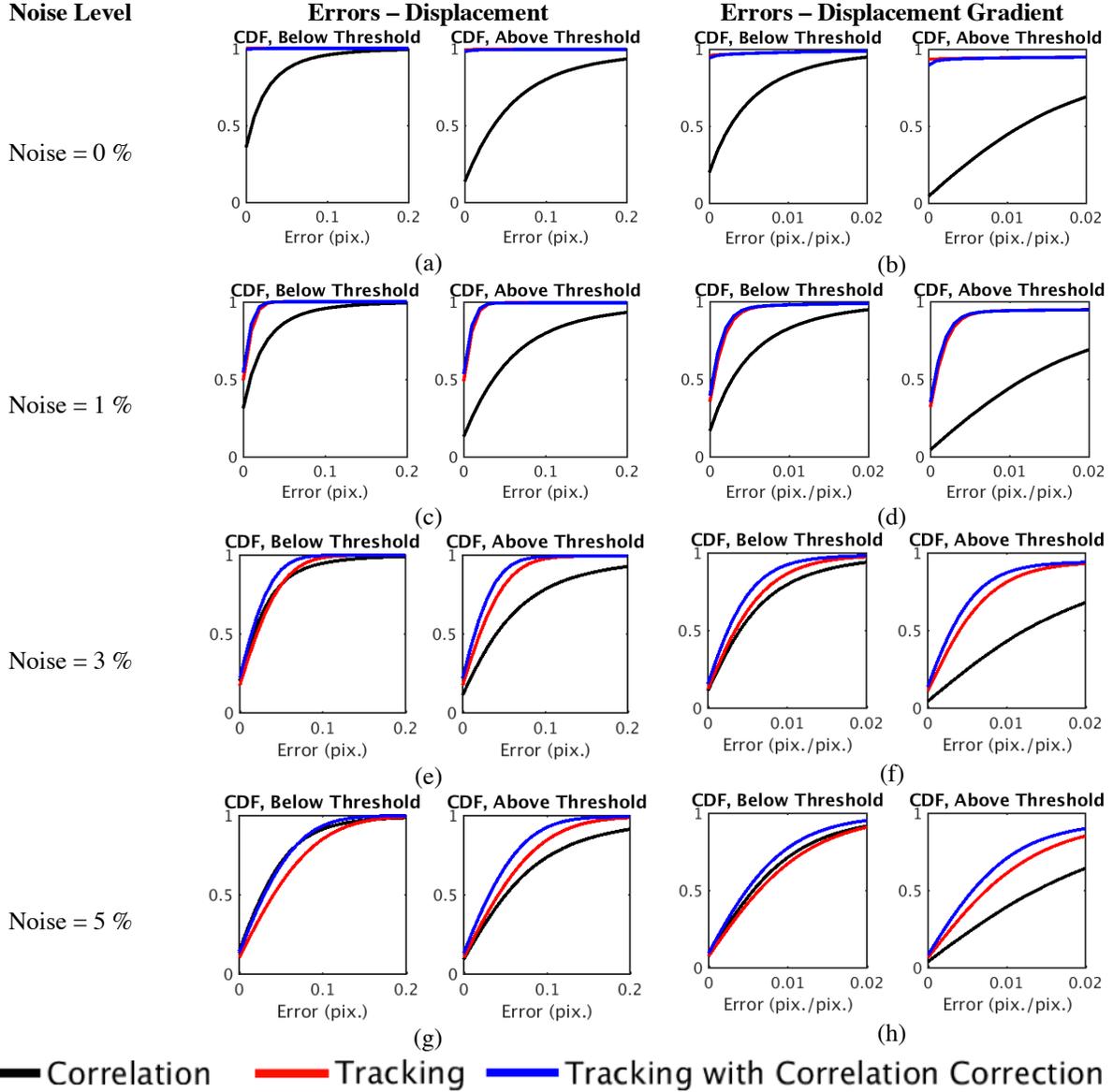

**Figure 5.** Error levels for the displacement and displacement gradient estimates obtained by the correlation and tracking methods.

The results of this analysis using synthetic images of physically realistic flow fields demonstrate that the proposed tracking approach, with non-overlapping dots, apriori identification and correlation correction provides a significant reduction in error compared to the conventional cross-correlation method as well as a large improvement in the spatial resolution for flow fields with sharp density gradients.

## 5 Application to experimental images of flow exiting a converging-diverging nozzle

The tracking methodology was also applied to visualize the exit plane of a converging diverging nozzle for various pressure ratios. This flow field was chosen because of the presence of shocks, expansion fans and other interesting small scale features that appear at high pressure ratios, and serve as a good assessment of the spatial resolution offered by the algorithms. The nozzle geometry along with the experimental layout and a sample image of the target, is shown in Figure 6. A regular grid of dots was printed on a transparency and back-illuminated with an LED to serve as the dot pattern. The dots were 0.15 mm in diameter and had a spacing of 0.15 mm, designed to provide a dot diameter of 3-4 pixels to improve the subpixel position estimation, and a dot spacing of about 3-4 pixels to have about 15 dots in a 32x32 window for high spatial resolution. The chamber pressure was varied from 0 to 60 psi in steps of 5 psi, while the exit pressure was maintained at atmospheric conditions (14.7 psi). For each pressure condition, the flow was allowed to reach steady state before capturing the images. The images were recorded using a PCO Pixelfly camera and a zoom lens set at a focal length of 105 mm.

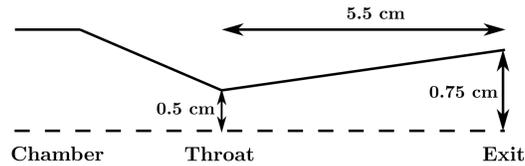

(a) Nozzle Geometry

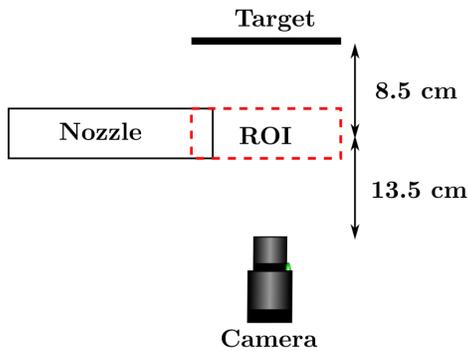

(b) Experimental Layout (Top View)

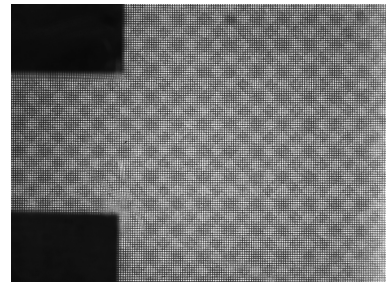

(c) Sample image of the BOS dot pattern

**Figure 6**. Details of the experimental setup used to visualize the flow in the exit plane of a converging-diverging nozzle.

The images of the dot pattern with and without the flow were analyzed using the tracking and correlation methods described before, and the displacement contours for two chamber pressures are shown in Figure 7.

From the displacement fields it can be seen that the results from the tracking analysis shown in Figure 7 (c)-(f) better capture small-scale features of the flow as compared to the correlation results shown in Figure 7 (a)-(b), which appear highly smoothed. Further it is seen that the tracking with correlation correction, shown in Figure 7 (e)-(f) provides a smoothing of the noisy displacement field compared to (c)-(d) while maintaining the high spatial resolution. The increase in spatial resolution is also evident from the line plots shown in (g)-(h) where tracking is able to better capture the sharp jumps in the displacement field. Overall, the dot tracking methodology is seen to be reliable when applied to BOS experimental data using dot patterns of high dot densities, while also increasing the spatial resolution of the measurements.

# 6 Conclusions

In this paper we proposed a dot tracking methodology for processing BOS images with high dot density based on two features of BOS experiments: (1) low displacements (2-3 pixels) and (2) availability of prior information about dot locations and size from manufacturing. We use the prior information about the dot locations to perform dot identification and sizing without the need for an intensity threshold, making the method more robust to image noise. We also proposed an improvement to the final displacement estimation, where we correlate the intensity maps of the matched dots instead of subtracting their centroid locations, to improve the performance in high noise situations. In this way we are able to combine the high spatial resolution benefit of tracking with the noise robustness property of correlation methods.

We analyzed the performance of this method and compared it to the conventional cross-correlation algorithm using synthetic and experimental BOS images. For synthetic BOS images of buoyancy driven turbulence, the tracking methods far outperformed the correlation method especially with low image noise and in regions with a requirement for high spatial resolution. For higher noise levels the errors in the tracking algorithms increased due to the position error from the subpixel fit being sensitive to image noise; however the tracking method with the correlation correction at the end was robust to this effect as the final displacement estimation does not depend on the centroid estimation process, and performed best overall.

For experimental BOS images of the flow field in the exit of a converging-diverging nozzle, the tracking methods again performed the best, by being able to resolve sharp changes in the density field in the presence of shocks and expansion fans.

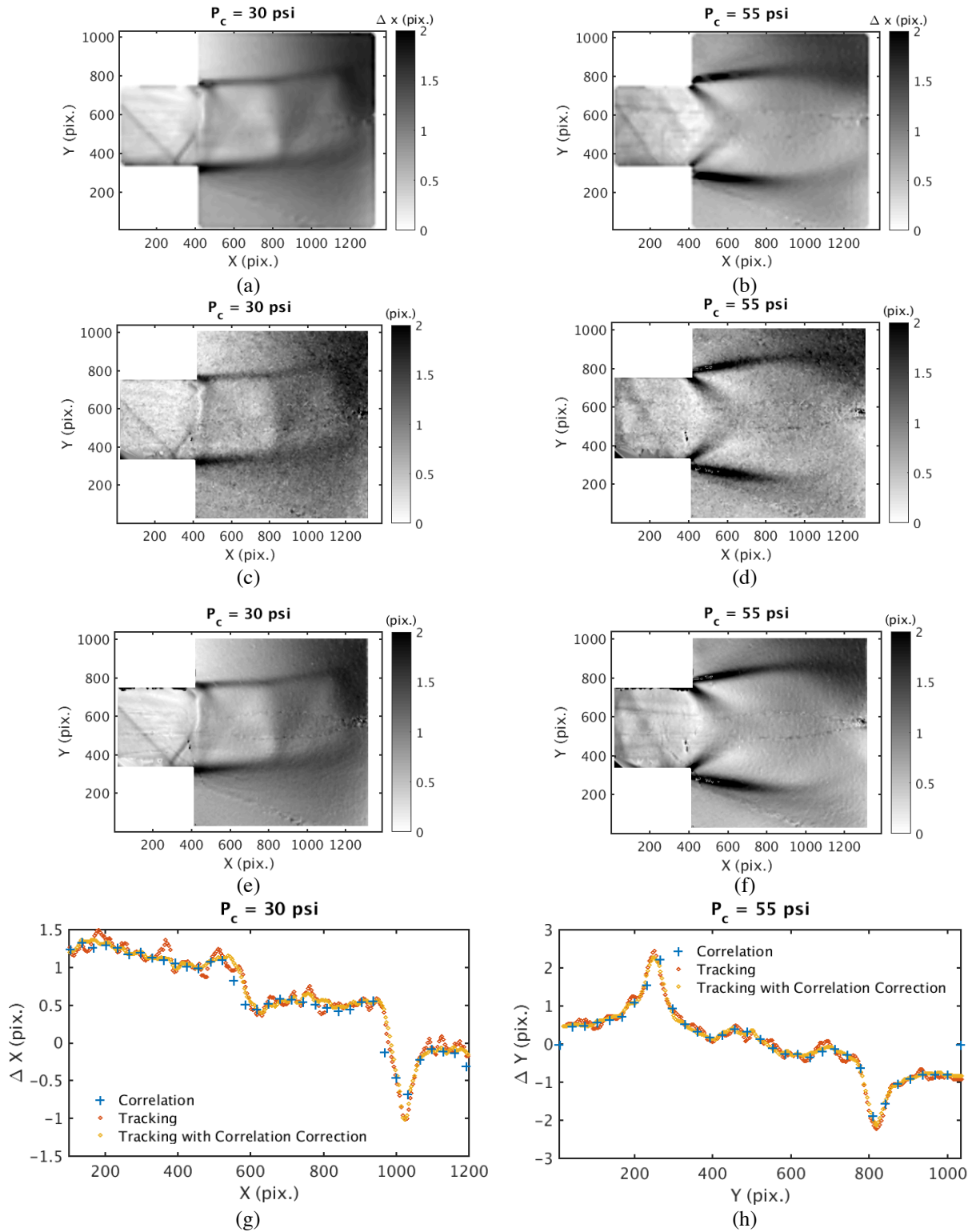

**Figure 7**. Flow in the exit plane of a converging-diverging nozzle, obtained from the different processing methods. Left column is for a chamber pressure of 30 psi, and right column is for 55 psi. (a)-(b) Correlation, (c)-(d) Tracking without Correlation Correction, (e)-(f) Tracking with Correlation Correction, (g) Line plot of displacements along X for Y = 600 pix., (h) Line plot of displacements along Y for X = 500 pix.

# 7 Acknowledgment

Ravichandra Jagannath is acknowledged for help with the nozzle experiments. This material is based upon work supported by the U.S. Department of Energy, Office of Science, Office of Fusion Energy Sciences under Award Number DE-SC0018156.

# 8 References

[1]     M. Raffel, "Background-oriented schlieren (BOS) techniques," *Exp. Fluids*, vol. 56, no. 3, pp. 1–17, 2015.

[2]     G. Meier, "Computerized background-oriented schlieren," *Exp. Fluids*, vol. 33, no. 1, pp. 181–187, 2002.

[3]     S. B. Dalziel, G. O. Hughes, and B. R. Sutherland, "Whole-field density measurements by 'synthetic schlieren,'" *Exp. Fluids*, vol. 28, no. 4, pp. 322–335, 2000.

[4]     H. Richard and M. Raffel, "Principle and applications of the background oriented schlieren (BOS) method," *Meas. Sci. Technol.*, vol. 12, no. 9, pp. 1576–1585, 2001.

[5]     B. Atcheson, W. Heidrich, and I. Ihrke, "An evaluation of optical flow algorithms for background oriented schlieren imaging," *Exp. Fluids*, 2009.

[6]     M. J. Hargather and G. S. Settles, "A comparison of three quantitative schlieren techniques," *Opt. Lasers Eng.*, vol. 50, no. 1, pp. 8–17, Jan. 2012.

[7]     G. E. Elsinga, B. W. Van Oudheusden, F. Scarano, and D. W. Watt, "Assessment and application of quantitative schlieren methods: Calibrated color schlieren and background oriented schlieren," *Exp. Fluids*, vol. 36, no. 2, pp. 309–325, Feb. 2004.

[8]     R. Keane and R. Adrian, "Theory of cross-correlation analysis of PIV images," *Appl. Sci. Res.*, pp. 191–215, 1992.

[9]     J. Westerweel, "Fundamentals of digital particle image velocimetry," *Meas. Sci. Technol.*, vol. 8, no. 12, pp. 1379–1392, 1997.

[10]    F. Scarano and M. L. Riethmuller, "Iterative multigrid approach in PIV image processing with discrete window offset," *Exp. Fluids*, vol. 26, no. 6, pp. 513–523, 1999.

[11]    F. Scarano, "Iterative image deformation methods in PIV," *Meas. Sci. Technol.*, vol. 13, no. 1, pp. R1–R19, 2002.

[12]    F. Scarano and M. L. Riethmuller, "Advances in iterative multigrid PIV image processing," *Exp. Fluids*, vol. 29, no. 7, pp. S051–S060, 2000.

[13]    M. Marxen, P. E. Sullivan, M. R. Loewen, B. J. Èhne, and B. Jähne, "Comparison of Gaussian particle center estimators and the achievable measurement density for particle tracking velocimetry," *Exp. Fluids*, vol. 29, no. 2, pp. 145–153, Aug. 2000.

[14]    N. D. Cardwell, P. P. Vlachos, and K. A. Thole, "A multi-parametric particle-pairing algorithm for particle tracking in single and multiphase flows," *Meas. Sci. Technol.*, vol. 22, no. 10, p. 105406, Oct. 2011.

[15]    Y. G. Guezennec and N. Kiritsis, "Statistical investigation of errors in particle image velocimetry," *Exp.*

*Fluids*, vol. 10, no. 2–3, pp. 138–146, Dec. 1990.

[16] F. Charruault, A. Greidanus, and J. Westerweel, "A dot tracking algorithm to measure free surface deformations," *Exp. Fluids*, 2018 (submitted).

[17] K. Ohmi and H.-Y. Li, "Particle-tracking velocimetry with new algorithms," *Meas. Sci. Technol.*, vol. 11, no. 6, pp. 603–616, Jun. 2000.

[18] M. R. Brady, S. G. Raben, and P. P. Vlachos, "Methods for Digital Particle Image Sizing (DPIS): Comparisons and improvements," *Flow Meas. Instrum.*, vol. 20, no. 6, pp. 207–219, 2009.

[19] S. M. Soloff, R. J. Adrian, and Z.-C. Liu, "Distortion compensation for generalized stereoscopic particle image velocimetry," *Meas. Sci. Technol.*, vol. 8, no. 12, pp. 1441–1454, 1997.

[20] M. Raffel, C. E. Willert, S. T. Wereley, and J. Kompenhans, *Particle image velocimetry: a practical guide*. Springer, 2013.

[21] D. Livescu, C. Canada, K. Kalin, R. Burns, I. Staff, and Pulido, "Homogeneous Buoyancy Driven Turbulence Data Set," no. 1, pp. 1–7, 2014.

[22] D. LIVESCU and J. R. RISTORCELLI, "Buoyancy-driven variable-density turbulence," *J. Fluid Mech.*, vol. 591, pp. 43–71, 2007.

[23] Y. Li, E. Perlman, M. Wan, Y. Yang, C. Meneveau, R. Burns, S. Chen, A. Szalay, and G. Eyink, "A public turbulence database cluster and applications to study Lagrangian evolution of velocity increments in turbulence," *J. Turbul.*, vol. 9, no. December 2016, p. N31, 2008.

[24] E. Perlman, R. Burns, Y. Li, and C. Meneveau, "Data exploration of turbulence simulations using a database cluster," *Proc. 2007 ACM/IEEE Conf. Supercomput. (SC '07)*, 2007.

[25] L. K. Rajendran, S. P. M. Bane, and P. P. Vlachos, "PIV/BOS Synthetic Image Generation in Variable Density Environments for Error Analysis and Experiment Design," Dec. 2018.

[26] A. Sharma, D. V. Kumar, and A. K. Ghatak, "Tracing rays through graded-index media: a new method.," *Appl. Opt.*, vol. 21, no. 6, pp. 984–987, 1982.

[27] C. Brownlee, V. Pegoraro, S. Shankar, P. McCormick, and C. Hansen, "Physically-based interactive schlieren flow visualization," *IEEE Pacific Vis. Symp. 2010, PacificVis 2010 - Proc.*, pp. 145–152, 2010.

[28] M. Born and E. Wolf, *Principles of optics*, 6th ed. Pergamon Press, 1980.

[29] M. Raffel, C. E. Willert, S. T. Wereley, and J. Kompenhans, *Particle Image Velocimetry*. Berlin, Heidelberg: Springer Berlin Heidelberg, 2007.

[30] L. K. Rajendran, S. P. M. Bane, and P. P. Vlachos, "PIV/BOS Synthetic Image Generation in Variable Density Environments for Error Analysis and Experiment Design," Dec. 2018.

[31] L. Venkatakrishnan and G. E. a. Meier, "Density measurements using the Background Oriented Schlieren technique," *Exp. Fluids*, vol. 37, no. 2, pp. 237–247, 2004.